\documentstyle[12pt]{article}
\newcommand{\beq}{\begin{equation}}
\newcommand{\eeq}{\end{equation}}
\begin{document}

\begin{titlepage}
\begin{center}
{\hbox to\hsize{hep-th/9905221  \hfill   MIT-CTP-2860}} 

{\hbox to\hsize{ 
\hfill PUPT-1860}}

{\hbox to\hsize{ 
\hfill BUHEP-99-9}}

\bigskip
\vspace{3\baselineskip}

{\Large \bf  A Large Mass Hierarchy
 
from a Small Extra  Dimension }

\bigskip

\bigskip

{\bf  Lisa Randall}\\
\smallskip

{ \small \it  
Joseph Henry Laboratories,
Princeton University,
Princeton, NJ 08543, USA\\

and

 Center for Theoretical Physics,

Massachusetts Institute of Technology, Cambridge, MA 02139, USA }

\smallskip
{\bf Raman Sundrum}\\
\smallskip
{\small \it Department of Physics, 

Boston University, Boston, MA 02215, USA}

\bigskip

{\tt  randall@baxter.mit.edu} \\
{\tt sundrum@budoe.bu.edu}

\bigskip

\vspace*{.5cm}

{\bf Abstract}\\
\end{center}
\noindent
We propose a new higher-dimensional mechanism for solving the hierarchy 
problem. The weak scale is generated from a large scale
of order the Planck scale through an exponential hierarchy.
However, this exponential arises not from gauge interactions
but from the background metric (which is a slice of 
$AdS_5$ spacetime). 
This mechanism relies on the existence of only a single 
additional dimension. We demonstrate a simple  explicit example
of this mechanism with two three-branes, one of which contains the Standard
Model fields. The experimental
consequences of this scenario are new and dramatic.
There are fundamental spin-2 excitations with mass 
of weak scale order, which are  coupled with {\it weak scale}
as opposed to gravitational strength  
to the standard model particles. The phenomenology
of these models is  quite distinct from that
of  large extra dimension scenarios; none of the current
constraints on theories with  very large extra dimensions apply.

\bigskip

\bigskip

\end{titlepage}

\section{Introduction}

If spacetime is fundamentally higher dimensional
with  $4 + n$ spacetime dimensions, then the effective four-dimensional 
(reduced) Planck scale, $M_{Pl}= 2 \times 10^{18}$ 
GeV,  is determined by the fundamental 
$(4+n)$-dimensional 
Planck scale, $M$, and the geometry of the extra dimensions. In the simplest 
cases, where the higher dimensional spacetime is approximately a product of a 
4-dimensional spacetime with a $n$-dimensional compact space, 
\begin{equation}
\label{product}
M_{Pl}^2 = M^{n+2} V_{n},
\end{equation}
where $V_n$ is the volume of the compact space.  
Recently it has been proposed that the large 
hierarchy between the weak scale and the 
fundamental scale of gravity can be eliminated by taking the compact space to 
 be very large \cite{large}. 
The fact that we do not see experimental signs of 
the extra dimensions despite the fact that the compactification scale, 
$\mu_c \sim 1/V_{n}^{1/n}$, 
 would have to be much smaller than the weak scale,
implies that  the SM particles and forces with the exception of gravity 
are confined to a $4$-dimensional 
subspace within the 
$(4 + n)$-dimensional spacetime, referred to as a ``3-brane''. 
 While this scenario does eliminate the hierarchy between the
weak scale $v$ and the Planck scale $M_{Pl}$, 
it 
introduces a new hierarchy, namely that between $\mu_c$ and $v$. In light of 
this it is worthwhile to explore alternatives.

Here we will present a distinct higher dimensional scenario
which provides an alternative approach to generating the
hierarchy. We propose that the metric is not factorizable,
but rather the four-dimensional metric is multiplied
by a ``warp'' factor which is a rapidly changing function of an
additional dimension. 
 The dramatic consequences for the hierarchy problem that
we identify in this letter 
follow from the particular non-factorizable metric,
\begin{equation}
\label{solution1}
ds^2 = e^{- 2 k r_c \phi} \eta_{\mu \nu} dx^{\mu} dx^{\nu} + r_c^2 d \phi^2, 
\end{equation}
where $k$ is a scale of order the Planck scale, 
 $x^{\mu}$ are coordinates for the familiar four 
dimensions, while $0 \leq \phi \leq \pi$ 
is the coordinate for an extra dimension, which is a 
finite interval whose size is set by $r_c$. 
We will show that this  metric is a solution to
Einstein's equations in a simple set-up with two 3-branes
and appropriate cosmological terms.
In this space, four-dimensional 
mass scales are related to five-dimensional input mass parameters
 and the warp factor, $e^{- 2 k r_c \phi}$.  To
generate a large hierarchy does not require extremely
large $r_c$. This is because the source of the hierarchy
is an {\it exponential} function of the compactification radius.
 The small exponential factor above  is  the source of the large 
hierarchy between the observed Planck and weak scales.

Although designed to address the hierarchy problem 
by exploiting an additional  dimension,
this  solution  is quite distinct from that proposed in Ref. \cite{large}:
1) The hierarchy between the fundamental five-dimensional Planck scale and 
the compactification scale, $\mu_c \equiv 1/r_c$ is only of order 50, as 
opposed to $(M_{Pl}/TeV)^{2/n}$. 2) There
is one additional dimension, as opposed to $n \geq 2$.
The experimentally distinctive consequences are:
1) There are no light Kaluza-Klein modes. The excitation scale
is of order  a TeV. Therefore, current constraints from
particle physics \cite{collider}, astrophysics  
and cosmology \cite{astro} do not apply. 
Because of this the scale at which gravity becomes strong 
 can be quite low. However, as with the scenario of Ref. \cite{large},  
string/M-theoretic excitations are also expected to appear at the TeV scale.
2) The coupling of an individual KK excitation to
matter or to other gravitational modes is set
by the weak, not the Planck scale. The KK modes are {\it not}
invisible; they should be observable at high
energy colliders as spin-2 resonances that
 can be reconstructed from their decay products.

\section{The Set-Up}

Because our spacetime does not fill out all of five dimensions,
we need to specify boundary conditions, which we take to be periodicity
in $\phi$, the angular coordinate parameterizing the fifth dimension,
supplemented with the identification of $(x, \phi)$ with $(x, -\phi)$;
that is we work on the space $S^1/{\bf Z_2}$. We take the range of $\phi$
to be from $-\pi$ to $\pi$; however the metic
is completely specified by the values in the range $0 \leq \phi \leq \pi$.
The orbifold fixed points at $\phi = 0, \pi$ will be taken as the locations of
 two 3-branes, extending in the $x^{\mu}$-directions, so that they are the 
boundaries of the five-dimensional spacetime. The 3-branes 
can support 
$(3+1)$-dimensional field theories. Both couple to the purely four-dimensional 
components of the bulk metric:
\begin{equation}
\label{smmetric}
g_{\mu \nu}^{vis}(x^{\mu}) \equiv G_{\mu \nu}(x^{\mu}, \phi = \pi), \ \  \ 
g_{\mu \nu}^{hid}(x^{\mu}) \equiv G_{\mu \nu}(x^{\mu}, \phi = 0),
\end{equation}
where $G_{MN}$,  $M, N = \mu, \phi$, is the five-dimensional metric.

%The orbifold fixed points at $\phi = 0, \pi$ will be taken as the locations of

This set-up is in fact similar to the scenario of Ref. \cite{large}. However,
we take into account the effect of the branes on the bulk
gravitational metric. 
Working out the consequences of the localized
energy density peculiar to the brane set-up, we find a new solution to the 
hierarchy problem. As we will show, this requires nothing beyond the 
existence of the 3-branes in five dimensions and their compatibility with  
four-dimensional Poincare invariance.

The classical action describing  the above set-up is given by 
\begin{eqnarray}
\label{action}
S &=& S_{gravity} + S_{vis} + S_{hid} \nonumber \\
S_{gravity} &=& 
\int d^4 x \int_{- \pi}^{\pi} d \phi \sqrt{-G} \{- \Lambda + 2 M^3 R \} 
\nonumber \\
S_{vis} &=& \int d^4 x \sqrt{-g_{vis}} \{ {\cal L}_{vis} 
-  V_{vis} \} \nonumber \\
S_{hid} &=& \int d^4 x \sqrt{-g_{hid}} \{ {\cal L}_{hid} -  V_{hid} \}.
\end{eqnarray}
Note that from each 3-brane Lagrangian we have separated out a constant 
``vacuum energy'' which acts as a gravitational source 
even in the  absence of particle excitations. The detailed form of the rest of
 the 3-brane Lagrangians will not be relevant for determining the 
classical five-dimensional metric in the ground state. Further discussion of 
3-brane actions can be found in Ref. \cite{me1}.

\section{Classical Solution}

In this section we solve the five-dimensional Einstein's
equations for the above action,
\begin{eqnarray} \label{einstein}
\sqrt{-G} \left( R_{MN}-{1 \over 2 } G_{MN} R \right) &=& - \frac{1}{4 M^3} 
[ \Lambda \sqrt{-G} ~G_{MN} +  V_{vis} \sqrt{-g_{vis}} ~g_{\mu \nu}^{vis} 
~\delta^\mu_M \delta^\nu_N ~\delta(\phi - \pi) \nonumber \\ 
&+& 
 V_{hid} \sqrt{-g_{hid}}  ~g_{\mu \nu}^{hid} 
~\delta^\mu_M \delta^\nu_N ~\delta(\phi) ].
\end{eqnarray}

We assume there exists a solution that
 respects  {\it four}-dimensional 
Poincare invariance in the $x^{\mu}$-directions.
A five-dimensional metric satisfying this ansatz takes 
 the form 
\begin{equation}
ds^2 = e^{- 2 \sigma(\phi)} \eta_{\mu \nu} dx^{\mu} dx^{\nu} + r_c^2 d \phi^2. 
\end{equation}
The coefficient, $r_c$, is independent of $\phi$ when
 we have chosen coordinates such that
 $\phi$ is proportional to the proper 
distance in the extra direction, $r_c$ being the constant of proportionality. 
In this way, $r_c$ is the ``compactification radius'' of the extra dimensional
 circle prior to orbifolding. After orbifolding, the size of the extra 
dimensional interval is $\pi r_c$. $r_c$ is independent of $x$ by
 four-dimensional Poincare  invariance.

With this ansatz, the 
 Einstein's equations  following from Eq. (\ref{einstein}) reduce to 
\begin{equation}
\label{eom2}
\frac{6 \sigma^{\prime 2}}{r_c^2} = { - {\Lambda} \over 4 M^3}, 
\end{equation}
\begin{equation}
\label{eom1}
\frac{3 \sigma^{\prime \prime}}{r_c^2}  =  
\frac{{V}_{hid}}{4 M^3 r_c} \delta(\phi) +
\frac{{V}_{vis}} {4 M^3 r_c} \delta(\phi - \pi). 
\end{equation}

The solution to Eq. (\ref{eom2})  consistent with
the orbifold symmetry $\phi \to -\phi$ is 
\begin{equation}
\label{sigsol}
\sigma =  r_c |\phi| \sqrt{\frac{- {\Lambda}}{24M^3}}.
\end{equation}
The additive integration constant has been omitted because it just amounts 
to an overall constant rescaling of the $x^{\mu}$. Clearly this 
solution only makes sense if $\Lambda < 0$,  which we will therefore 
assume to be the case from now on.
Note that the spacetime in between the two 3-branes is simply a slice 
of an $AdS_5$ geometry.\footnote{Note, this makes our bulk 
gravitational dynamics compatible with a supersymmetric extension.}

Recall that in computing derivatives we are to consider the metric a periodic 
function in $\phi$. Eq. (\ref{sigsol}), valid for $-\pi \leq \phi \leq \pi$, 
 then implies 
\begin{equation}
\sigma^{\prime \prime} = 2 r_c \sqrt{\frac{- {\Lambda}}{24M^3}}
[\delta(\phi) - \delta(\phi - \pi)]. 
\end{equation}
{}From this we see that we only obtain a solution to Eq. (\ref{eom1}) 
if $V_{hid}, V_{vis}, \Lambda$ are related in terms of a single scale
$k$,   
\begin{equation}
\label{conditions}
V_{hid} = - V_{vis} = 24 M^3 k, ~~ \Lambda = - 24 M^3 k^2.
\end{equation} 
 These relations between the boundary and bulk
cosmological terms
are required in order to obtain
 a solution that respects four-dimensional
Poincare invariance. 
 Note that   precisely these 
relations arise in the five-dimensional effective theory of the 
Horava-Witten scenario
if one were to interpret the expectation values of
the background 3-form field (but with frozen Calabi-Yau moduli) 
as cosmological
terms in the effective five dimensional
theory after Calabi-Yau compactification \cite{hw1}.  
We will assume that $k <
 M$ so that the bulk curvature is  small compared to the higher dimensional 
Planck scale and we trust our solution.  

Our final solution for the bulk metric  is then
\begin{equation}
\label{solution}
ds^2 = e^{- 2 k r_c |\phi|} \eta_{\mu \nu} dx^{\mu} dx^{\nu} + r_c^2 d \phi^2.
\end{equation}
The compactification radius
 $r_c$ is effectively an arbitrary integration constant for this solution.

\section{Physical Implications}

We are considering  a  small  $r_c$ (but still larger
than ${1/k}$). Therefore,   the fifth dimension
cannot be resolved in present (or future) gravity experiments;  
spacetime appears  
four-dimensional. It therefore makes sense to use a four-dimensional
effective field theory description.  In this
section, we derive the parameters of this low-energy theory,
namely the four-dimensional Planck scale, and the mass parameters
of the four-dimensional fields, in terms of the five-dimensional scales, $M, 
k, r_c$. 

 The first 
step is to identify  the massless gravitational fluctuations about our 
classical solution, Eq. (\ref{solution}). These will provide the gravitational
 fields for our effective theory. They are  the 
zero-modes of our classical solution, and take the form 
\begin{equation}
\label{zeromodes}
ds^2 = e^{- 2 k T(x) |\phi|} [\eta_{\mu \nu}+ \overline{h}_{\mu \nu}(x)]
 dx^{\mu} dx^{\nu} + 
T^2(x) d \phi^2. 
\end{equation}
Here, $\bar{h}_{\mu \nu}$ represents tensor 
fluctuations about Minkowski
space and  is the physical graviton of the four-dimensional
effective theory (and is the massless mode in the Kaluza-Klein decomposition 
of $G_{\mu \nu}$).
Note that this metric is {\it locally} the same as our ``vacuum'' solution, 
Eq. (\ref{solution}), since any smooth four-dimensional metric, 
\begin{equation}
\overline{g}_{\mu \nu}(x) \equiv
\eta_{\mu \nu}+\overline{h}_{\mu \nu}(x),
\end{equation}
is locally Minkowskian, while any smooth real 
function $T(x)$ is locally constant. The compactification radius, $r_c$, is
 the vacuum expectation value of the modulus field, $T(x)$. 
As with many higher dimensional theories, it will be critical
 that  the $T$ modulus is stabilized at its
vacuum expectation value $r_c$, with a mass of at least $10^{-4}$ eV.
Although an essential element of the theory, this problem
is not yet solved
(but see Refs. \cite{me2}). From now on, we replace   $T$ with $r_c$.
In compactifying 
extra dimensions, 
one frequently encounters vector zero modes from $A_{\mu} d 
x^{\mu} d \phi$ fluctuations of the metric 
(that is the original Kaluza-Klein idea), corresponding to 
the continuous isometries of the higher dimensions, but in the present case 
there are no such isometries in the presence of the 3-branes. So all such 
off-diagonal 
fluctuations of the metric are massive and excluded from the low-energy 
effective theory.

The four-dimensional effective theory now follows by substituting 
Eq. (\ref{zeromodes}) 
 into the original action, Eq. (\ref{action}). 
We focus on the curvature term from which we can
derive the scale of gravitational interactions: 
%\begin{eqnarray}
%S_{eff} &=&  \int d^4 x \int_{- \pi}^{\pi} d \phi 
%\sqrt{-\overline{g}} \{ - 2 \Lambda 
%r_c e^{-4 k T |\phi|} - 12 M^3 T \overline{g}^{\mu \nu} \partial_{\mu}
%e^{-k T(x) |\phi|}  \partial_{\nu} e^{-k T(x) |\phi|} +
%2 M^3 T(x) e^{-2 k T |\phi|}  \overline{R} \} \nonumber \\
%&+& \int d^4 x \sqrt{-\overline{g}}   \{ V_{vis} e^{- 4 k r_c \pi} - 
%V_{hidden}
%\},
%\end{eqnarray}
\begin{equation}
 S_{eff} \supset \int d^4 x \int^{\pi}_{-\pi} d \phi ~
2 M^3 r_c e^{-2 k r_c  |\phi|}  \sqrt{-\overline{g}} ~\overline{R}
\end{equation}
where $\overline{R}$ denotes the four-dimensional Ricci scalar
 made out of 
$\overline{g}_{\mu \nu}(x)$, in contrast to the five-dimensional Ricci 
scalar, $R$, made out of $G_{MN}(x, \phi)$. 
Because the low-energy fluctuations do not change the $\phi$ dependence (the 
effective fields depend on $x$ alone), we can 
explicitly perform the $\phi$ integral to obtain a purely four-dimensional 
action. From this we derive
\begin{equation}
\label{effplanck}
M_{Pl}^2  =  M^3 r_c \int_{-\pi}^{\pi} d\phi  e^{-2 k r_c |\phi|} = 
\frac{M^3}{k}
[1 - e^{- 2 k r_c \pi}].
\end{equation}
This is an important result. It tells us
that $M_{Pl}$ depends only weakly on $r_c$ in
the large $k r_c$ limit.
Although the exponential has very little effect
in determining  the Planck scale,
we will now see that it plays a crucial role in the
determination of the visible sector masses.

In order to determine the matter field Lagrangian we need to know the 
 coupling of  the 3-brane fields  to the low-energy 
gravitational fields, in particular the metric, $\overline{g}_{\mu \nu}(x)$. 
 From Eq. (\ref{smmetric}) we see that $g_{hid} = 
\overline{g}_{\mu \nu}$. 
This is not the case for the visible sector
 fields;   by Eq. (\ref{smmetric}), we have 
$g^{vis}_{\mu \nu} = e^{- 2 k r_c \pi} \overline{g}_{\mu \nu}$. By properly 
normalizing the fields we can determine the physical masses. Consider for 
example a fundamental Higgs field,
\begin{equation}
S_{vis} \supset \int d^4 x \sqrt{-g_{vis}} \{ g_{vis}^{\mu \nu} D_{\mu}
 H^{\dagger} D_{\nu} H - \lambda (|H|^2 - v_0^2)^2 \}, 
\end{equation}
which contains one mass parameter $v_0$. Substituting Eq. 
(\ref{smmetric}) into this action yields 
\begin{equation}
S_{vis} \supset \int d^4 x \sqrt{- \overline{g}} e^{- 4 k r_c \pi} 
\{ \overline{g}^{\mu \nu} e^{2 k r_c \pi} D_{\mu}
 H^{\dagger} D_{\nu} H - \lambda (|H|^2 - v_0^2)^2 \}, 
\end{equation}
After wave-function renormalization, $H \rightarrow e^{k r_c \pi} H$, we 
obtain
\begin{equation}
S_{eff} \supset \int d^4 x \sqrt{- \overline{g}}  
\{ \overline{g}^{\mu \nu} D_{\mu}
 H^{\dagger} D_{\nu} H - \lambda (|H|^2 - e^{-2 k r_c \pi} v_0^2)^2 \}. 
\end{equation}
A remarkable thing has happened. We see that the physical mass scales are set 
by a symmetry-breaking scale,
\begin{equation}
\label{weak}
v \equiv e^{- k r_c \pi} v_0.
\end{equation}
This result is completely general: any mass parameter 
$m_0$ on the visible 3-brane 
in the fundamental higher-dimensional theory will correspond to a 
physical mass
\begin{equation}
\label{punch}
m \equiv e^{- k r_c \pi} m_0
\end{equation}
when measured with the metric $\overline{g}_{\mu \nu}$, which is the metric 
that appears in the effective Einstein action, since all operators
get rescaled according to their four-dimensional conformal weight.
  If 
$e^{k r_c \pi}$ is of order $10^{15}$, this mechanism  produces 
TeV physical mass scales from fundamental mass parameters not far from the 
Planck scale, $10^{19}$ GeV.  Because this geometric factor is an 
exponential, we clearly do not require very large hierarchies among the 
fundamental parameters, $v_0, 
k, M,$ and  $\mu_c \equiv 
1/r_c$; in fact, we only require $k r_c \approx 50$.

Having established the relevant masses for matter fields,
we turn to the question of the gravitational modes
themselves. This gives rise
to a rich and very distinctive phenomenology. 
To determine the parameters of the
gravitational modes  in  detail, requires an explicit Kaluza-Klein
decomposition. We will do this in Ref. \cite{next}. The
result is that the masses {\it and} couplings of
the Kaluza-Klein modes are determined by the 
TeV scale. This result can be readily understood.

Until this point, we have viewed $M \approx M_{Pl}$ 
as the fundamental scale, and the TeV scale
as a derived scale as a consequence of the
exponential factor appearing in the metric.
However, one could equally well have regarded
the TeV scale as fundamental,
and the Planck scale of $10^{19}$ GeV as
the derived scale. That is, the ratio
is the physical dimensionless quantity.
{}From this viewpoint, which is the one naturally
taken by a four-dimensional observer residing
on the visible brane, the large Planck scale
(the weakness of gravity)  arises
because of the small overlap of the graviton
wave function in the fifth dimension (which is the warp factor) with our brane.
In fact, this is the {\it only} small number produced.
All other scales are set by the TeV scale.

Technically, this change in viewpoint is established
by the change of coordinates, $x^\mu \to e^{k r_c \pi} x^\mu$.
In this case, the warp factor at $\phi=\pi$ is unity,
whereas that at $\phi=0$ is $e^{2 k r_c} \pi$. 
In this language, since there is no rescaling of the
``$v$'' parameter in the Higgs potential because 
the Higgs is already canonically normalized, 
the scale $v$ should take  its physical value. Because
we are assuming all fundamental mass parameters
are of the same order, all these parameters are also  of order 
TeV \footnote{Note
that the relation between the  mass  parameters in
the new coordinates and the old mass parameters  is
due to the spacetime coordinate rescaling.}. 

%However, one finds 

This result 
contrasts sharply with the scenario of large extra dimensions for solving the 
hierarchy problem with a product structure for the full spacetime, where the 
Kaluza-Klein splittings are much smaller than the weak scale, possibly smaller 
than an eV. The dangerous astrophysical and cosmological effects of very light
Kaluza-Klein states are absent in our model.

The phenomenological implications of this scenario
for future collider searches are very distinctive.
 For a product spacetime, each excited state 
couples with gravitational strength, and the key to observing these states in 
accelerator experiments is the large multiplicity of states due to their fine 
splittings. In our model, with roughly weak scale splittings a relatively small
number of excitations will be kinematically accessible at accelerators.
However their couplings to matter are set by the weak scale rather than the 
Planck scale. 
Instead of gravitational strength couplings $\sim {\rm Energy}/M_{Pl}$, each
excited state coupling is of order $
 {\rm Energy}/{\rm TeV}$, 
and therefore each can be {\it individually} detected. These 
resonances can be detected via their decay products. This should allow detailed
reconstruction, permitting mass and spin determination of these {\it 
gravitational} modes.

{}From the above discussion it should be
 clear that at energies somewhat larger than 
the weak scale, the excited gravitons are strongly coupled. This  
regime should likely open up the production of string/M-theoretic 
excitations which lie outside the domain of even 
our starting five-dimensional field theory.
This means that although the fundamental scales of the higher dimensional
theory are of order $M_{Pl}$, the {\it apparent} 
 scale where the theory becomes strongly coupled
and the string/M excitations appear is of order the weak
scale according to a four-dimensional observer.
This is an important result for the consistency of our scenario
beyond tree level. As with Ref. \cite{large}, the TeV-scale strings will cut 
off large
renormalization of the weak scale.

\section{Conclusions} 

In the 3-brane scenario, where extra-dimensional translational symmetry is 
necessarily broken, non-trivial warp factors naturally arise upon 
solving Einstein's equations. The Kaluza-Klein 
reduction is considerably more subtle than in product spacetimes, as we will 
detail in a following paper \cite{next}.
This has important phenomenological and theoretical implications.

 In this letter, we focussed on  a potential 
phenomenological implication of this scenario,
namely an exponential generation of the hierarchy.
 Remarkably, the four-dimensional 
masses on the visible brane depend on the background metric in such
a way that their physical values differ significantly
from the input mass parameters, even without invoking
a large compactification volume. This is a potential 
resolution to the hierarchy problem akin in spirit
to the ideas of strongly coupled gauge theories which
generate the low scale through an exponential times
a fundamental high energy scale. 
As an aside,
we mention that the exponential we  exploited could generate
other scales, such as the low-energy supersymmetry
breaking scale.
However,  it is important to the viability of
our mechanism that it is possible
to stabilize the compactification radius
roughly two orders 
 of magnitude 
larger than the fundamental five dimensional Planck length. 
Issues such as flavor violation and proton decay
in the face of the low scale of new physics \cite{flavor}, 
also remain important challenges.

Fortunately, this solution to the hierarchy problem 
is subject to experimental
verification. The phenomenology is quite distinct from
the scenario of large radius compactification. 
The gravitational resonances are of order a TeV,
and couple with TeV suppressed, rather than Planck-suppressed,
strength. Furthermore, there are no
experimental bounds pushing this scale
very high.  Should this solution prove correct, there
is a rich spectroscopy awaiting us  at the LHC.

{\bf Acknowledgements:} We wish to acknowledge
useful conversations  
with Ignatios Antoniadis, Nima Arkani-Hamed, Micha Berkooz, 
Alvaro deRujula, Savas Dimopoulos, 
Martin Gremm, Igor Klebanov, Gary
Horowitz, Albion Lawrence, Takeo Moroi, Burt Ovrut, Joe Polchinski,
Andy Strominger, Herman Verlinde, and Dan Waldram.

\end{document}